\documentclass[lettersize,journal]{IEEEtran}
\usepackage{amsmath,amsfonts,amssymb}
\usepackage{algorithmic}
\usepackage{algorithm}
\usepackage{array}
\usepackage[caption=false,font=normalsize,labelfont=sf,textfont=sf]{subfig}
\usepackage{textcomp}
\usepackage{stfloats}
\usepackage{url}
\usepackage{verbatim}
\usepackage{graphicx}
\usepackage{cite}
\usepackage{float}
\usepackage{multirow}
\usepackage{booktabs}
\begin{document}

\title{Cochleagram-based Noise Adapted Speaker Identification System for Distorted Speech}

% Affiliations
\author{\IEEEauthorblockN{Sabbir Ahmed, Nursadul Mamun, Md Azad Hossain}\\
\IEEEauthorblockA{\textit{Department of Electronics and Telecommunication Engineering} \\
\textit{Chittagong University of Engineering and Technology}\\
Chattogram, Bangladesh \\
sabbirahmed.cuet.ete@gmail.com, (nursad.mamun, azad)@cuet.ac.bd}
}



\maketitle

\begin{abstract}
Speaker Identification refers to the process of identifying a person using one's voice from a collection of known speakers. Environmental noise, reverberation and distortion make the task of automatic speaker identification challenging as extracted features get degraded thus affecting the performance of the speaker identification (SID) system. This paper proposes a robust noise adapted SID system under noisy, mismatched, reverberated and distorted environments. This method utilizes an auditory features called cochleagram to extract speaker characteristics and thus identify the speaker. A $128$ channel gammatone filterbank with a frequency range from $50$ to $8000$ Hz was used to generate 2-D cochleagrams. Wideband as well as narrowband noises were used along with clean speech to obtain noisy cochleagrams at various levels of signal to noise ratio (SNR). Both clean and noisy cochleagrams of only $-5$ dB SNR were then fed into a convolutional neural network (CNN) to build a speaker model in order to perform SID which is referred as noise adapted speaker model (NASM). The NASM was trained using a certain noise and then was evaluated using clean and various types of noises. Moreover, the robustness of the proposed system was tested using reverberated as well as distorted test data. Performance of the proposed system showed a measurable accuracy improvement over existing neurogram based SID system.
\end{abstract}


\section{Introduction}
\IEEEPARstart{S}{peaker} identification is a biometric process that determine the speaker's identity analyzing underlying speech data. The task of SID can be done using two types of speech datasets - text dependent and text independent. In text dependent speaker identification system, a speaker needs to utter some specific words, phrases or sentences whereas text independent speaker identification system does not require utterance of any specific word or sentence to identify a speaker. In this study, text dependent as well as text independent datasets were used. Automatic SID has a wide range of applications such as accessing database, voice mail, providing security for confidential information, voice dialing, forensics and so on. However, unlike many other biometrics such as fingerprint, face and iris, human voice/speech is a performance based biometric as speaker's characteristic information is embedded in the way of speaking. This means that speech signal is highly variable. Stationary, impulsive or time varying environmental background noise \cite{rose1994integrated}, room reverberation \cite{greenberg2010human} and distortion are some of the few external variability factors that contribute to mismatched speaker identification. In this paper, the performance degradation obeserved in the existing SID systems due to speech signals affected by any of the three variability factors is considered. In order to confront the dilapidation of performance, the idiosyncratic characteristics of speaker also known as features should have large inter-speaker variability, small intra-speaker variability, the capability of retaining speaker's underlying characteristics for noisy and distorted speech signal.

A large number of speech features has been proposed over the last few decades to perform speaker identification under varied background conditions. Almost all of the feature extraction methods of speech can be categorized into two broad categories - voice/speech production system modeling and peripheral auditory system modeling. Linear prediction cepstral coefficients (LPCC) derived from linear prediction providing an array of cepstral coefficients is one of the features that is based on the human voice production model \cite{makhoul1976correction}.
LPCC models the human the voice production system and extracts speech formants using an all-pole filter. This feature works well in clean condition. However, significant performance degradation observed due to a large spectral distortion in the spectral envelope under noisy environments \cite{li2001auditory}.
Based on transforms, features based on auditory periphery can be again divided into two types - Fourier transform (FT) based and auditory transform based features. Mel frequency cepstral coefficients (MFCCs) is a FT based feature and a very widely used feature in speaker recognition. This feature uses fast Fourier transform (FFT) to generate a linear scale spectrum. Afterwards, a set of band-pass filters is applied on the FFT output along a mel scale \cite{davis1980comparison}. Although the MFCC features perform very well achieving up to $100\%$ accuracy by Nakagawa et. al in clean condition \cite{nakagawa2011speaker} but similar to the LPCC features it cannot perform well in mismatched conditions \cite{li2010auditory}. Another peripheral auditory feature based on FT called the perceptual linear predictive (PLP) analysis. To subdue the constant factors of the spectral component \cite{hermansky1994rasta}, relative spectra also called the RASTA is cascaded with the PLP feature to get the RASTA-PLP features \cite{hermansky1991rasta}. MFCC as well as RASTA-PLP use fast and efficient FFT algorithm but the time-frequency resolution of FT is not same as that of our hearing system. The reasons are it produces pitch-harmonics and the frequency bands in FT are distributed linearly unlike the distribution of human cochlea. Moreover, FFT spectrogram are more susceptible to noise and computational distortion compared to auditory transforms \cite{li2010auditory}. Efforts were made with the aim of extracting features without including noise from the characteristic information of speaker e.g. cepstrum mean normalization (CMS) \cite{furui1981cepstral}, warping \cite{pelecanos2001feature} and robust parameterization \cite{nikias1993signal}. However, these techniques were found to have a minimal effectiveness against additive distortions and non-linear channel effects \cite{islam2016robust}. 

Cochlear filter cepstral coefficients (CFCC) a new feature proposed by Li et al. in \cite{li2010auditory} is a auditory transform based feature which emulates human peripheral hearing mechanism. It showed an improved performance over the FT based features and also found to show robustness to noise in different training and testing conditions. Linear auditory filters and database dependent parameters used in this study were not as suitable for achieving a robustness for various corpora. Ganapathy et. el proposed a new feature called the frequency domain linear prediction (FDLP) coefficient in \cite{ganapathy2012feature}, uses a autoregressive model on the high energy peaks of the signal and was reported achieve $30\%$ accuracy improvement over MFCC feature. A substantial improvement in speaker identification under noisy environments has been reported by Zhao et. al in \cite{zhao2012casa} with the use of a new auditory based feature called Gammatone frequency cepstral coefficeints (GFCCs) by producing a binary time-frequency (T-F) mask employing a technique mentioned as computational auditory scene analysis (CASA).

Apart from modeling human speech production system and peripheral auditory system, Bosko et. al proposed two dimensional information entropy feature based on the amplitude-time trajectory which is considered to contain speaker specific characteristics \cite{bozilovic2015text}. However, this feature was only tested for text independent speaker identification and a small dataset consisting of only six speakers. Tjandra et. al utilized the combination of spectrogram as well as cochleagram features for automatic speaker recognition using both deep neural network (DNN) and convolutional neural network (CNN). A relative reduction in phoneme error rate up to $8.2\%$ for CNN and $19.7\%$ for DNN was reported in \cite{tjandra2015combination}. A novel feature based on the responses of the auditory nerve (AN) called neurogram has been proposed by Zilany et. al is capable of replicating the non-linear phenomena seen in the auditory periphery at different levels such as in the cochlea, inner hair cell (IHC), IHC-AN synapse and AN fibers \cite{zilany2014updated}. The non-linear phenomena includes phase response, two-tone suppression, non-linear tuning and several high level effects \cite{zilany2014updated, zilany2006modeling, zilany2009phenomenological}. Unlike the acoustic features, neural responses exhibits phase locking property making the feature robust under various background conditions \cite{miller1997effects}. However, for pink and white gaussian noise GFCC exhibited better performance in case of wideband and resulted higher accuracy at $0$dB SNR compared to the neurogram feature \cite{islam2016robust}. Besides, the neurogram feature also performed poorly for text independent datasets compared to the GFCC features. Recently, two new features have been proposed i-vector \cite{dehak2010front} and x-vector \cite{snyder2018x} which are extracted from the output of an inner layer of a DNN as speaker embedding but the DNN is fed with MFCC feature to obtain those feature vectors. The size of feature vectors depend on the number of neurons in that inner layer. Ravenelli et. al proposed a raw speech waveform based method of speaker identification using a DNN called SincNet \cite{ravanelli2018speaker} but this method was evaluated only for clean speech.

For dereverberation, CMN is the most naive approach \cite{wang2005robust, wang2007robust}. However, impulse response in far distance talking environment is much longer than the analysis window for short term spectral analysis. So, CMN is suitable for late reverberation. Several research have been conducted to address this issue \cite{jin2007far, kinoshita2009suppression, wang2012dereverberation, kinoshita2006spectral, wang2013hands}. Beamforming is also a simple and robust approach for spatial filtering \cite{van1988beamforming, hughes1999performance}. Therefore, it is suitable for dereverberation as it can be used to restrain any signal from noise or the direction of reflection \cite{wang2012dereverberation, seltzer2004likelihood}. In 2011, Habets et. al proposed two stage beamforming for noise reduction as well as dereverberation \cite{habets2011joint} though neither of the approaches could achieve good performance for very strong reverberation \cite{zhang2015deep}. Nakatani et. al proposed unified and maximum likelihood convolutional beamforming techniques to denoise and dereverberate speech simultaneously which performs better than the previous other beamformer methods \cite{nakatani2019unified, nakatani2019maximum}. Neural networks are also now being used for dereverberation in speech/speaker recognition because of their flexible representations \cite{nugraha2014single}. Denoising autoencoders (DAEs) are now being used in speech processing along with image processing to denoise speech signal. DAE showed effectiveness in noise reduction as well as dereverberation as this model is capable of extracting only the important features to map in the latent space \cite{vincent2010stacked, lu2013speech}. Trimula et. al proposed deep autoencoder for speaker identification for noisy speech \cite{tirumala2017deep}. However such model contains many layers which computationally expensive. Speech separation and dereverberation using a two stage multimodal network was proposed by Tan et. al \cite{tan2020audio} and speech enhancement using DAE with multi-branched encoder has been proposed by Yu et. el \cite{yu2020speech}. However, these methods utilize very complex and deep neural network having multiple branches as well as the output of such models were not evaluated for speaker identification to realize how much speaker specific information they could retain.

In this study, the challenge of speaker identification under noisy, reverberated and distorted condition was addressed using a noise adapted CNN. The network was trained using a mixture of both clean and noisy 2-D cochleagrams to adapt the network to noisy data. However, the speaker model was evaluated using $9$ other different noises at $5$ different level of SNR including the adapted noise for both text dependent and text independent datasets. Besides, the model was also evaluated for reverberated, noisy reverberated and distorted data. Finally, the proposed system was compared with the existing noise robust feature e.g. neurogram and analyzed the performance for reverberated and mismatched conditions. The performance of the proposed system showed a substantial improvement at low SNR level for all noises.
The rest of this paper is organized as follows: Methodology of the proposed speaker identification system in section II. Section III presents the experimental results and evaluations followed by discussion in section IV. Finally conclusions and future works are presented in section V.

% METHODOLOGY
\section{Methodology}
\begin{figure}[H]
    \centering
    \includegraphics[width=0.48\textwidth]{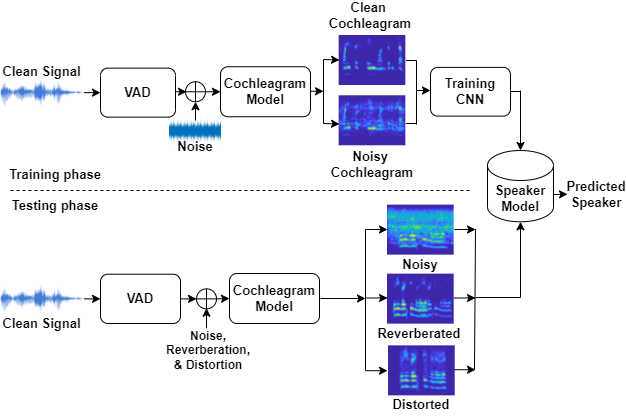}
    \caption{ Block diagram of the proposed method of speaker identification system }
    \label{fig:1}
\end{figure}

The proposed methodology used for performing speaker identification is illustrated in this section. In the training phase, the clean speech samples and the noisy speech samples obtained by addition of $10$ different noises such as - airport, babble, car, exhibition, pink, restaurant, station, street, train and white gaussian having SNR = $-5$ dB were mixed to obtain $10$ different datasets. All those datasets were used to train the CNN classifier and in testing phase the best adapted model was used to evaluate the model on noisy, reverberated and distorted test samples which is depicted in Fig.\ref{fig:1}. Each of the step in the proposed method is briefly described in the following subsections.

\subsection{Datasets}
In this study, one text dependent and two text independent corpora were used to train and evaluate the model.
\subsubsection{Text Dependent}
This study used 'University Malaya (UM)' dataset for text dependent speaker identification. It consists of $39$ speakers having $10$ samples for each speaker where each speaker utters `university malaya' in each of the samples. The dataset was recorded in the University of Malaya in a sound proof room at $8$kHz sampling rate \cite{islam2015neural}.

\subsubsection{Text Independent}
\begin{itemize}
\item YOHO Dataset:\\
It is a text independent dataset for speaker identification consisting of $138$ speakers. This study used $129$ speakers each having $72$ utterances whose analog bandwidth is $3.8$kHz. The sampling frequency of the speech samples is $8$kHz.

\item NIST Speaker Recognition Evaluation (SRE) Dataset:\\
NIST SRE dataset was developed by Linguistic Data Consortium (LDC) and National Institute of Standards and Technology (NIST). There are about $2,255$ hours of telephone speech (American). The sampling frequency is $16$kHz. There are $100$ speakers each having $100$ speech samples.
\end{itemize}
\subsection{Preprocessing}
Silent and low energy parts of speech contain very little or no information about a speaker. So, these portions are unnecessary and can be removed using a voice activity detector (VAD). VAD removes the unwanted parts from speech and concatenates the fragmented parts. In order to align the fragmented components dynamic time warping (DTW) can be used. However, in this study DTW was not used rather a fixed size 2D cochleagrams were used. Finally, the sampling frequency of the speech samples(for SRE dataset) were downsampled to $8$ kHz to reduce the computational cost.

\subsection{Addition of Noise and Reverberation}
After removal of the unwanted parts from the speech samples, $10$ different types of noises (9 narrowband and 1 wideband) were added with the speech samples. The clean speech signals were modulated with noises at SNR $-5$dB, $0$dB, $5$dB, $10$dB and $15$dB. However, this study used a noisy signal with $-5$dB SNR along with the clean signal to train the model.

Moreover, the impacts of reverberation were evaluated at time delays - $800$ms, $500$ms, $200$ms and $100$ms. Reverberation were added to clean as well as noisy speech samples. But for noisy data, reverberation was added to only white gaussian and babble noise mixed data which represent the effect of reverberation on wideband and narrowband noisy data respectively.

\subsection{Addition of Distortion}
Clipping which is a form of distortion was used in this study to represent distorted speech data. Clipping limits a signal when it exceeds a certain threshold. Two types of clipping namely - center clipping and peak clipping were added with the raw audio speech samples to generate distorted cochleagram. For an input signal t(n), the center clipped output signal z(n) can be expressed as follows:
\begin{equation}
z(n) = \begin{cases}
			(t(n) - C_{th}), & \text{$t(n)\geq C_{th}$}\\
			0, & \text{$\mid t(n) \mid < C_{th}$}\\
			(t(n) + C_{th}), & \text{$t(n)\leq -C_{th}$}
		\end{cases}
\end{equation}
where $C_{th}$ represents the clipping threshold. Usually, clipping threshold is $30\%$ of the highest magnitude of the signal. The maximum value of first and last $\frac{1}{3}$ of the signal are measured to get the high $C_{th}$. The lowest value is taken as the maximum magnitude and $60\%$ to $80\%$ of that maximum magnitude is set to be the $C_{th}$.
On the other hand, peak clipping clips or limits the peaks of a signal which may cause when the gain of an amplifier is increased too much.

\subsection{Cochleagram Model}
\begin{figure}[H]
\centering
\includegraphics[width=0.48\textwidth]{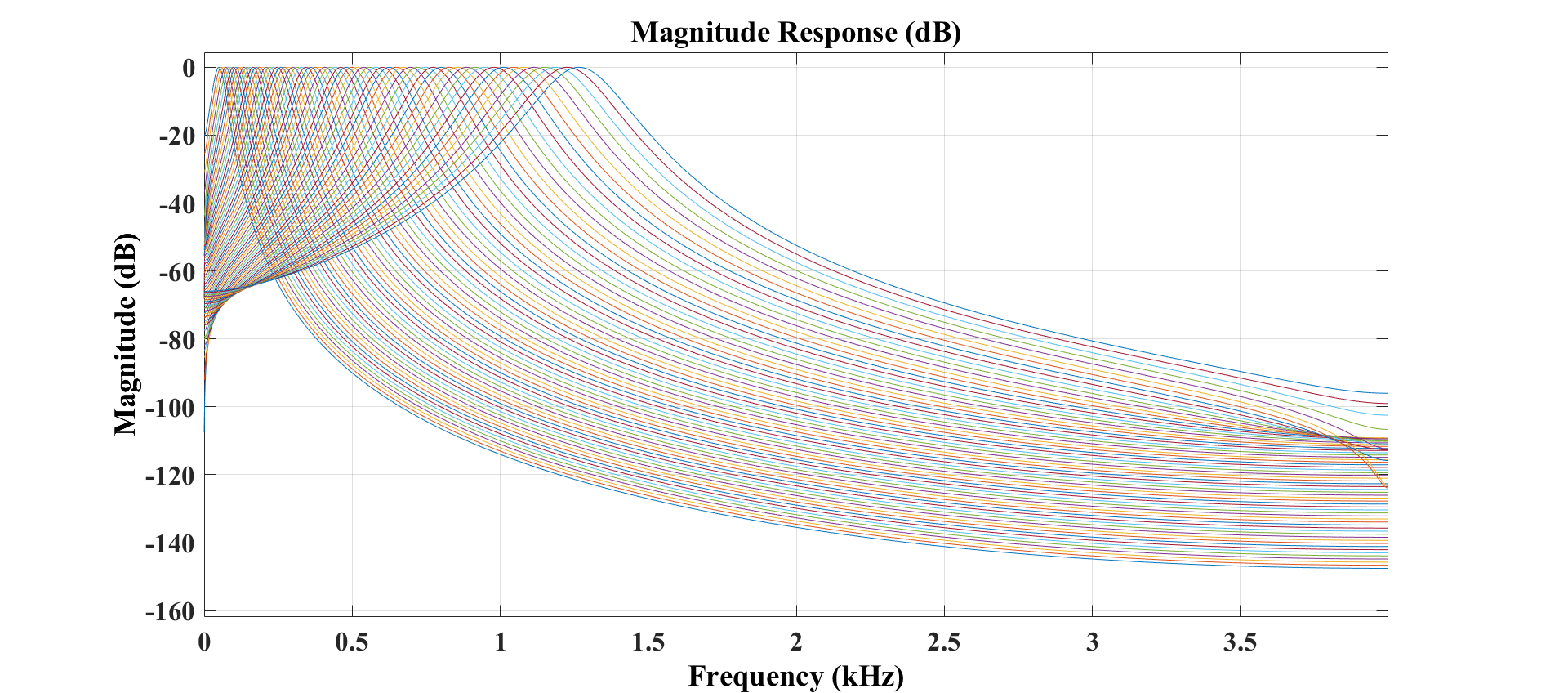}
\caption{Magnitude response of the gammatone filterbank}
\label{fig:2}
\end{figure}

In speaker identification, a suitable feature is a very important factor for robust speaker identification because most of the features are susceptible to noise. In this study we used time sequences of gammatone features (GFs) to generate cochleagrams which is a two dimensional T-F representation of speech signal. At first the input speech signal is decomposed into the time frequency domain by auditory filtering using gammatone filterbank. The gammatone filterbank resembles the cochlear filtering and derived from the auditory periphery. For cochleagram generation, 128 channel gammatone filterbank with each channel having $50\%$ overlap between consecutive frames was used over $50$ to $8000$Hz frequency range. The cochleagram model generates cochleagrams of size $875 \times 656$.  If the center frequency of a gammtaone filter is$f$, its impulse response is given by -

\begin{equation}
h(f,t) = \begin{cases}
			t^{n-1}e^{-2\pi wt}\cos(2\pi ft), & \text{$t\geq1$}.\\
			0, & \text{else}.
		\end{cases}
\end{equation}
where $n$ represents the order of filters, $w$ is the rectangular bandwidth. If $f_c$ is the center frequency, the output response $y(m,t)$ can be written as - 
\begin{equation}
y(m, t) = y(t) * h(f_c,t)
\end{equation}
where ``$*$'' represents convolution in time domain and $m$ represents a particular filter channel. To compensate the filter delay, the filter response is delayed by $(n-1) /(2\pi b)$. Fig.\ref{fig:2} shows the magnitude response of the gammatone filterbank for sampling frequency $f_s = 8$kHz.

\subsection{Model Architecture and Training}
\begin{figure}[H]
\centering
\includegraphics[width=0.48\textwidth]{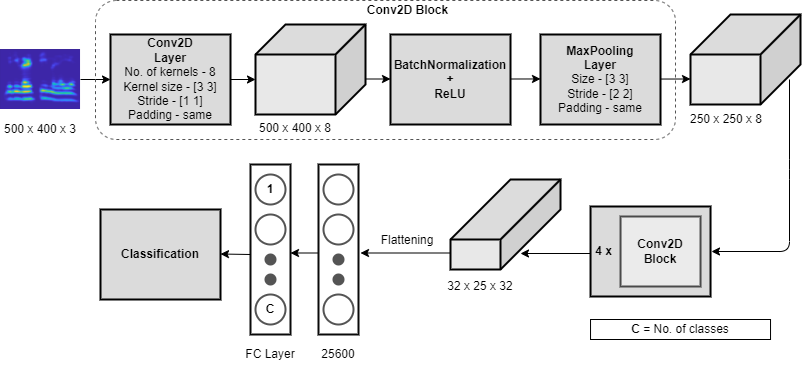}
\caption{ Architecture of the convolutional neural network }
\label{fig:3}
\end{figure}

For speaker identification, a convolutional neural network (CNN) was used which consists of five convolutional layers and a dense layer as the output layer. Each of the convolutional layer uses rectified linear unit (ReLU) as activation followed by a BatchNormalization layer and a MaxPooling layer as illustrated in fig. \ref{fig:3}. Each of the first two convolutional layers use $8$ kernels, the third, fourth and fifth layers use $16$, $32$ and $32$ kernels respectively each of size $3 \times 3$. 
The kernels were initialized using `Glorot' initializer. The convolutional kernels uses stride $1 \times 1$ with padding `same'. The kernel size in the MaxPooling layer is the same as that of the convolutional layer with a stride of $2 \times 2$. As optimizer, Stochastic Gradient Descent with Momentum (SGDM) was used with a learning rate of $lr = 1e^{-3}$. $L_2$ regularization was used on the weights and biases of each of the convolutional layer to reduce overfitting. The model was trained for $30$ epochs for a minibatch size of $64$. Besides, for reducing the computational cost of neural network, the cochleagrams were resized to $500 \times 400$ before feeding into the network.

% RESULT ANALYSIS
\section{Experimental Results and Evaluations}
The performance of the proposed noise adapted SID model was evaluated using noisy, clean reverberated and noisy reverberated and distorted data. As mentioned in the earlier section, $10$ different models were adapted using $10$ different noises. However, from all the $10$ different models, one model that performed best on all the different noisy data was selected to evaluate all three datasets.
\subsection{Performance of NASM Under Noisy Conditions}
\begin{figure}[H]
\centering
\includegraphics[width=0.49\textwidth]{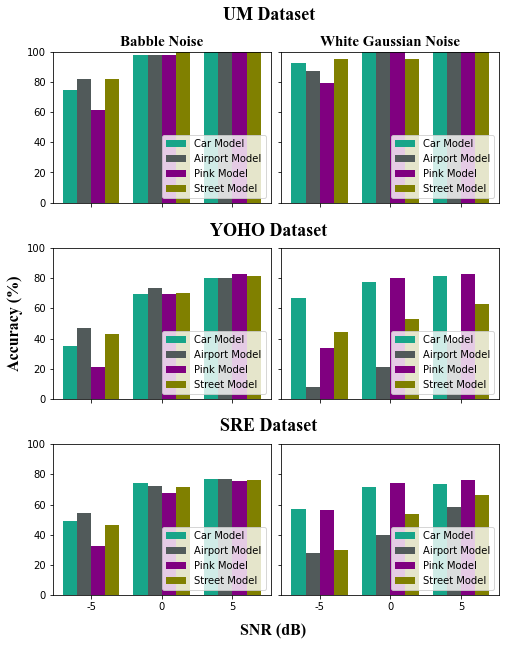}
\caption{Performance comparison among airport, car, pink and street noise adapted models tested using babble and white gaussian noise. }
\label{fig:4}
\end{figure}
\begin{figure}[H]
    \centering
    \includegraphics[width=0.4\textwidth]{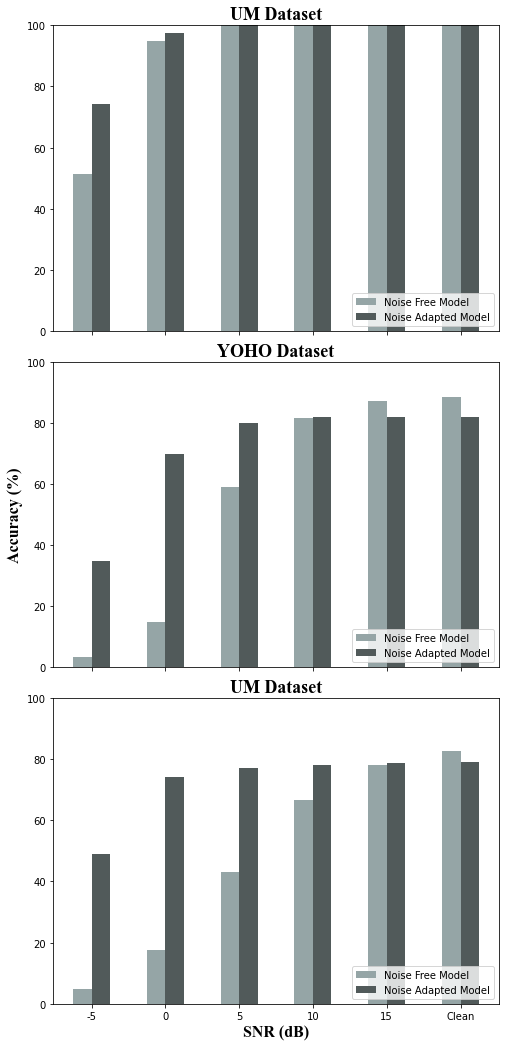}
    \caption{Performance comparison between noise free model and noise adapted model tested on babble noise}
    \label{fig:5}
\end{figure}

The best noise adapted model found for all the three corpora is the network trained with clean and car noise added data. As shown in fig. \ref{fig:4} babble noise which is a narrowband noise and white gaussian noise which is a wideband noise were considered for the comparison to provide a notion about the performance on both types of noises. In the training data, car noise was added at $-5$dB with clean data, though the model was tested at $6$ different levels of SNR. To demonstrate the performance improvement in noisy environments, we compared the accuracy obtained from noise free model and noise adapted model. The comparison in \ref{fig:5} clearly delineates the noise adapted model's large performance improvement over noise free model in noisy conditions. However, performance on both the text independent datasets shows that the accuracy of the network at $15$dB and clean condition is slightly less than the noise free model. The performance at $15$dB and clean condition was found to increase $10\sim15\%$ in a separate experiment depending on the dataset when more data were used to train the network. For the same dataset used in this study, the model was trained for several configurations and architectures, yet the highest accuracy at $15$dB and clean condition did not show any considerable improvement. This suggests that the model is unable to converge more as the data for each of the sample largely differ from each other due to addition of noise at very low level of SNR representing data from two different sources for the same speech sample.

\begin{figure}[H]
    \centering
    \includegraphics[width=0.48\textwidth, height=17.5cm,keepaspectratio]{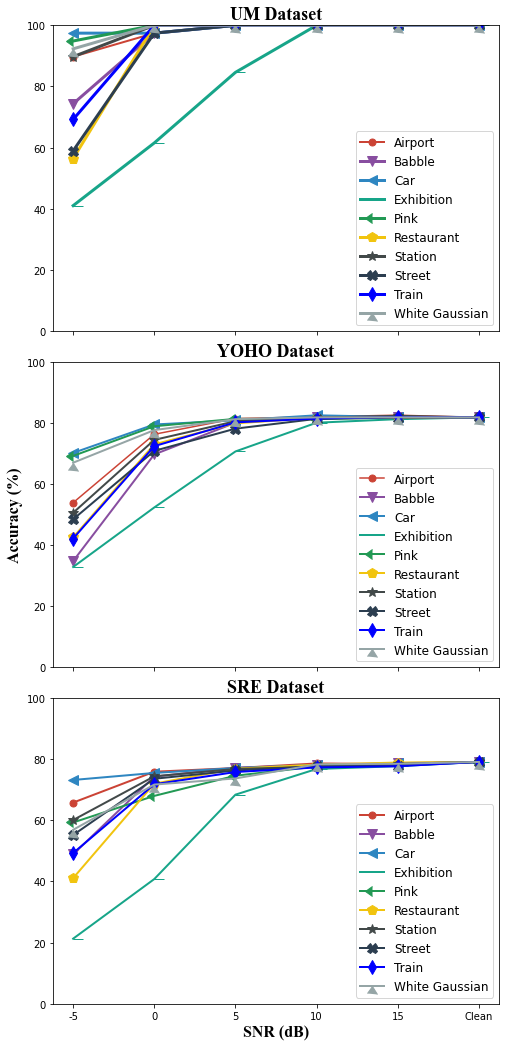}
    \caption{Performance of the car noise adapted model on various noises}
    \label{fig:6}
\end{figure}

One of the objectives of this study was to build a robust model that provides a consistent performance in mismatched conditions. To demonstrate the proposed model's noise robustness, the model was evaluated using $9$ other different noises along with the test set of same noise added data used to adapt the model for all the three datasets at various levels of SNR. Fig. \ref{fig:6} shows the accuracy obtained from the model for different noises. 

Performance on all the three different datasets shows that the performance improvement of the model obtained by training it using only one noise is reflected on other noises too. Another notable thing to mention is that white gaussian noise is a wideband noise, but the noise adapted model was trained using car noise which is a narrowband noise yet the accuracy for white gaussian noise did not get affected rather it is quite good.

\begin{figure}[H]
    \centering
    \includegraphics[width=0.48\textwidth]{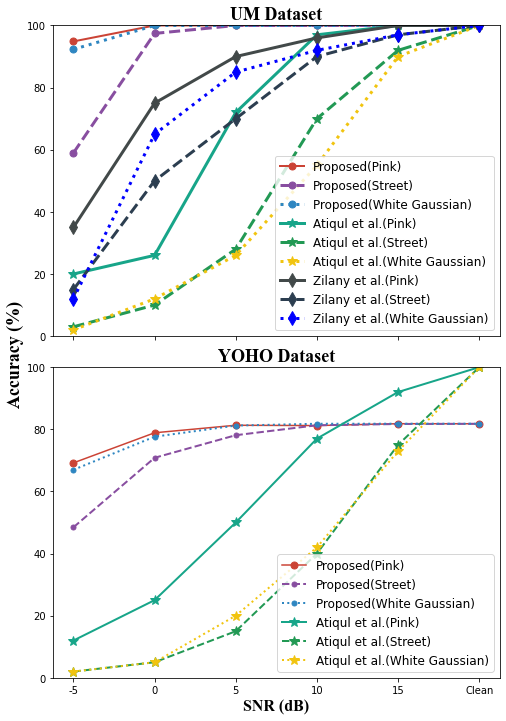}
    \caption{Performance comparison among the proposed cochleagram-based SID system and the neurogram-based SID systems proposed by Atiqul and Zilany et al.}
    \label{fig:7}
\end{figure}

After that, another comparison was made among the proposed model and the models based on neurogram features proposed by Atiqul et. al and Zilany et. al in \cite{islam2016robust, zilany2018novel}. The accuracy was compared for only pink, street and white gaussian noises of which both the pink and street noises are of narrowband type. The result in Fig. \ref{fig:7} depicts a large performance boost of the SID models for text dependent as well as text independent datasets particularly when the SNR level is low. Since the comparison was made using narrowband as well as wideband noise, it can be concluded that the proposed model will also exceed the accuracy of neurogram based models for other various noises.

\subsection{Performance of NASM Under Reverberant Conditions}
\begin{figure}[H]
    \centering
    \includegraphics[width=0.48\textwidth,height=17.5cm,keepaspectratio]{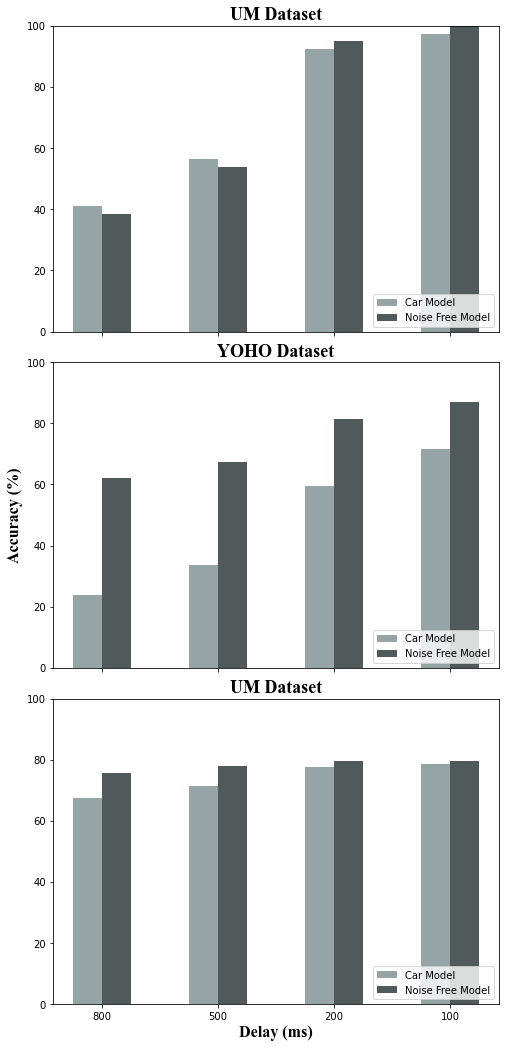}
    \caption{Accuracy of the noise free and the car noise adapted models on reverberated data.}
    \label{fig:8}
\end{figure}

The same car noise adapted model was tested by providing both reverberated data and noise with reverberation added data. Four different impulse delay times were introduced such as $800$ms, $500$ms, $200$ms and $100$ms to resemble reverberation from living room to large auditorium. The clean models were found to be more accurate than the noise adapted model as depicted in Fig. \ref{fig:8}.

\subsection{Performance of NASM Under Noisy and Reverberant Conditions}
When the performance was measured after addition of both reverberation and noise, the noise free model was found to perform much worse than the proposed noise adapted model particularly when the SNR is very low e.g. $-5$dB, $0$dB and $5$dB as shown in Table \ref{tab:my_label}. In order to test the models, babble and white gaussian noise were added with the reverberated data. 
\begin{table*}[t]
    \centering
    \caption{Performance of the car noise adapted model and noise free model on noisy and reverberated data}
    \begin{tabular}{c | c c c c c | c c c c c }
    \multicolumn{11}{c}{\textbf{UM Dataset (Car Model)}}\\
    \toprule
    \textbf{Noises}
    &  \multicolumn{5}{c|}{\textbf{Babble Noise}}
    &  \multicolumn{5}{c}{\textbf{White Noise}}\\
    \cmidrule(l){1-11}
    \textbf{Impulse Delay (ms)} & -5dB & 0dB & 5dB & 10dB & 15dB & -5dB & 0dB & 5dB & 10dB & 15dB\\
    \midrule
    800 & 20.51 & 28.21 & 35.90 & 35.90 & 41.03 & 25.64 & 30.77 & 38.46 & 38.46 & 41.03\\     
    500 & 38.46 & 48.72 & 48.72 & 56.41 & 58.97 & 41.03 & 48.72 & 46.15 & 51.28 & 58.97\\     
    200 & 76.92 & 87.18 & 97.44 & 92.31 & 92.31 & 74.36 & 84.62 & 94.87 & 89.74 & 92.31\\     
    100 & 66.67 & 92.31 & 94.87 & 97.44 & 97.44 & 89.74 & 97.44 & 92.31 & 97.44 & 97.44\\ 
    \toprule
    \end{tabular}
    
    \begin{tabular}{c | c c c c c | c c c c c }
    \multicolumn{11}{c}{\textbf{UM Dataset (Noise Free Model)}}\\
    \toprule
    \textbf{Noises}
    &  \multicolumn{5}{c|}{\textbf{Babble Noise}}
    &  \multicolumn{5}{c}{\textbf{White Noise}}\\
    \cmidrule(l){1-11}
    \textbf{Impulse Delay (ms)} & -5dB & 0dB & 5dB & 10dB & 15dB & -5dB & 0dB & 5dB & 10dB & 15dB\\
    \midrule
    800 & 5.13 & 17.95 & 41.03 & 35.90 & 41.03 & 23.08 & 17.95 & 30.77 & 35.90 & 35.90\\
    500 & 15.38 & 38.46 & 46.15 & 51.28 & 56.41 & 28.21 & 28.21 & 48.72 & 51.28 & 51.28\\ 
    200 & 25.64 & 76.92 & 87.18 & 94.87 & 94.87 & 53.85 & 69.23 & 94.87 & 92.31 & 94.87\\
    100 & 41.03 & 92.31 & 97.44 & 100.00 & 100 & 69.23 & 89.74 & 97.44 & 100.00 & 100.00\\
    \toprule
    \end{tabular}
    
    \begin{tabular}{c | c c c c c | c c c c c }
    \multicolumn{11}{c}{\textbf{YOHO Dataset (Car Model)}}\\
    \toprule
    \textbf{Noises}
    &  \multicolumn{5}{c|}{\textbf{Babble Noise}}
    &  \multicolumn{5}{c}{\textbf{White Noise}}\\
    \cmidrule(l){1-11}
    \textbf{Impulse Delay (ms)} & -5dB & 0dB & 5dB & 10dB & 15dB & -5dB & 0dB & 5dB & 10dB & 15dB\\
    \midrule
    800 & 5.14 & 11.24 & 17.34 & 22.29 & 23.55 & 15.79 & 20.35 & 22.77 & 22.87 & 24.13\\     
    500 & 7.66 & 16.76 & 26.16 & 31.49 & 33.33 & 22.67 & 28.78 & 31.88 & 32.75 & 33.14\\     
    200 & 15.02 & 36.53 & 51.26 & 57.66 & 58.82 & 42.44 & 52.62 & 56.88 & 58.14 & 59.50\\     
    100 & 20.35 & 51.26 & 65.50 & 69.00 & 71.41 & 53 & 66.18 & 69.09 & 70.54 & 71.22\\ 
    \toprule
    \end{tabular}
    
    \begin{tabular}{c | c c c c c | c c c c c }
    \multicolumn{11}{c}{\textbf{YOHO Dataset (Noise Free Model)}}\\
    \toprule
    \textbf{Noises}
    &  \multicolumn{5}{c|}{\textbf{Babble Noise}}
    &  \multicolumn{5}{c}{\textbf{White Noise}}\\
    \cmidrule(l){1-11}
    \textbf{Impulse Delay (ms)} & -5dB & 0dB & 5dB & 10dB & 15dB & -5dB & 0dB & 5dB & 10dB & 15dB\\
    \midrule
    800 & 2.33 & 8.33 & 25.19 & 41.18 & 56.01 & 1.65 & 5.33 & 13.57 & 29.17 & 47.87\\     
    500 & 2.62 & 7.66 & 30.52 & 48.84 & 61.24 & 1.74 & 5.52 & 15.21 & 33.33 & 54.36\\     
    200 & 2.71 & 12.89 & 46.61 & 67.93 & 78.59 & 2.23 & 9.30 & 24.13 & 49.90 & 73.26\\     
    100 & 3.29 & 14.24 & 55.23 & 77.23 & 85.27 & 3.20 & 10.27 & 27.52 & 55.81 & 79.65\\ 
    \toprule
    \end{tabular}
    
    \begin{tabular}{c | c c c c c | c c c c c }
    \multicolumn{11}{c}{\textbf{SRE Dataset (Car Model)}}\\
    \toprule
    \textbf{Noises}
    &  \multicolumn{5}{c|}{\textbf{Babble Noise}}
    &  \multicolumn{5}{c}{\textbf{White Noise}}\\
    \cmidrule(l){1-11}
    \textbf{Impulse Delay (ms)} & -5dB & 0dB & 5dB & 10dB & 15dB & -5dB & 0dB & 5dB & 10dB & 15dB\\
    \midrule
    800 & 27.10 & 55.80 & 62.30 & 65.10 & 66.70 & 36 & 54.00 & 57.90 & 64.20 & 65.90\\     
    500 & 30.00 & 60.60 & 68.20 & 70.40 & 71.10 & 40 & 58.50 & 63.10 & 68.00 & 71\\     
    200 & 34.90 & 68.10 & 74.4 & 75.8 & 76.90 & 46.70 & 64.00 & 68.8 & 73.90 & 76.60\\     
    100 & 35.00 & 71.50 & 77.20 & 77.70 & 78.20 & 49.70 & 65.70 & 70.4 & 75.50 & 78.50\\ 
    \toprule
    \end{tabular}
    
    \begin{tabular}{c | c c c c c | c c c c c }
    \multicolumn{11}{c}{\textbf{SRE Dataset (Noise Free Model)}}\\
    \toprule
    \textbf{Noises}
    &  \multicolumn{5}{c|}{\textbf{Babble Noise}}
    &  \multicolumn{5}{c}{\textbf{White Noise}}\\
    \cmidrule(l){1-11}
    \textbf{Impulse Delay (ms)} & -5dB & 0dB & 5dB & 10dB & 15dB & -5dB & 0dB & 5dB & 10dB & 15dB\\
    \midrule
    800 & 4.50 & 13.10 & 31.10 & 54.70 & 68.80 & 3.90 & 8.90 & 23.40 & 42.60 & 59.40\\     
    500 & 4.80 & 14.20 & 32.70 & 57.30 & 70.70 & 4.40 & 9.30 & 24.50 & 44.20 & 60.60\\     
    200 & 4.60 & 13.30 & 33.90 & 55.80 & 71.40 & 4.50 & 8.70 & 23.80 & 45.30 & 62.60\\     
    100 & 4.70 & 13.20 & 34.60 & 55.80 & 71.20 & 4.20 & 9.20 & 24.20 & 46.20 & 62.40\\ 
    \toprule
    \end{tabular}
    \label{tab:my_label}
\end{table*}
Table \ref{tab:my_label} shows that the performance of NASM and noise free model in case of text independent datasets varies largely under very low SNR. Moreover, accuracy at higher level of SNR was found to increase when the NASM was trained with noisy data of more than one SNR level. Thus the NASM model was found to be resistant to multiple distortion factors.

% \begin{figure}[H]
%     \centering
%     \includegraphics[width=0.48\textwidth, height=10.75cm]{figures/um_clean_vs_noise_adapted_when_noise_and_reverberation_added.png}
%     \caption{Performance of the car noise adapted model and the noise free model on noisy and reverberated UM dataset}
%     \label{fig:9}
% \end{figure}

% \begin{figure}[H]
%     \centering
%     \includegraphics[width=0.48\textwidth, height=10.75cm]{figures/yoho_clean_vs_noise_adapted_when_noise_and_reverberation_added.png}
%     \caption{Performance of the car noise adapted model and noise free model on noisy and reverberated YOHO dataset}
%     \label{fig:10}
% \end{figure}

% \begin{figure}[H]
%     \centering
%     \includegraphics[width=0.48\textwidth]{figures/sre_clean_vs_noise_adapted_when_noise_and_reverberation_added.png}
%     \caption{Performance of the car noise adapted model and the noise free model on noisy and reverberated SRE dataset}
%     \label{fig:11}
% \end{figure}

\subsection{Performance of NASM Under Distorted Conditions}
Two types of distortions - center clipping and peak clipping were introduced at three different threshold levels (30, 60 and 90) to the clean test set. Without noise, the car noise adapted model performed similar to that of the clean model. However, some of the other models adapted with various noises showed better performance than the noise free model as shown in Table \ref{tab:um_dataset}, \ref{tab:yoho_dataset} and \ref{tab:sre_dataset}. The models were tested $5$ times for each test set to observe the fluctuations in accuracy and the mean values were taken as the final result.
\begin{table}[H]
    \centering
    \caption{UM Dataset}
    \begin{tabular}{l | c c c | c c c|}
    \toprule
    \textbf{Thresholds}
    &  30ms & 60ms & 90ms & 30ms & 60ms & 90ms \\
    \cmidrule(l){1-7}
    \textbf{Models} 
    & \multicolumn{3}{c|}{\textbf{Center Clipping}} 
    & \multicolumn{3}{c|}{\textbf{Peak Clipping}} \\
    \midrule
    Airport & 100 & 97.44 & 89.74 & 79.49 & 92.31 & 97.44 \\
    Babble & 100 & 97.44 & 89.74 & 87.18 & 92.31 & 94.87 \\
    Exhibition & 100 & 97.44 & 94.87 & 94.87 & 97.44 & 100\\
    Pink & 100 & 97.44 & 89.74 & 82.05 & 89.74 & 92.31 \\
    Restaurant & 100 & 97.44 & 89.74 & 89.74 & 94.87 & 97.44 \\
    Station & 100 & 97.44 & 87.18 & 84.62 & 92.31 & 94.87\\
    Street & 100 & 97.44 & 89.74 & 92.31 & 92.31 & 97.44 \\
    Train & 100 & 97.44 & 92.31 & 87.18 & 92.31 & 94.87 \\
    White Gaussian & 97.44 & 94.87 & 84.62 & 87.18 & 87.18 & 89.74 \\
    Clean & 100 & 97.44 & 92.31 & 89.74 & 94.87 & 100 \\
    \end{tabular}
    \label{tab:um_dataset}
\end{table}
\begin{table}[H]
    \caption{YOHO Dataset}
    \begin{tabular}{l | c c c | c c c|}
    \toprule
    \textbf{Thresholds}
    &  30ms & 60ms & 90ms & 30ms & 60ms & 90ms \\
    \cmidrule(l){1-7}
    \textbf{Models} 
    & \multicolumn{3}{c|}{\textbf{Center Clipping}} 
    & \multicolumn{3}{c|}{\textbf{Peak Clipping}} \\
    \midrule
    Airport & 78.10 & 52.62 & 29.26 & 70.45 & 78.10 & 80.91 \\
    Babble & 78.20 & 50.68 & 26.45 & 72.09 & 79.26 & 82.07 \\
    Exhibition & 75.58 & 41.18 & 20.06 & 75.29 & 81.20 & 83.14\\
    Pink & 80.33 & 66.67 & 45.35 & 71.12 & 78.97 & 82.56 \\
    Restaurant & 75.19 & 36.92 & 16.76 & 72.19 & 80.62 & 82.85 \\
    Station & 78 & 47.67 & 25.87 & 69.38 & 79.17 & 82.07 \\
    Street & 78.39 & 52.33 & 28 & 74.61 & 81.20 & 83.62 \\
    Train & 76.84 & 50.29 & 26.07 & 72 & 79.26 & 81.20 \\
    White Gaussian & 78.29 & 61.92 & 42.93 & 72.77 & 80.91 & 82.85\\
    Clean & 77.42 & 46.32 & 25.48 & 76.07 & 84.69 & 87.40 \\
    \end{tabular}
    \label{tab:yoho_dataset}
\end{table}

\begin{table}[H]
    \caption{SRE Dataset}
    \begin{tabular}{l | c c c | c c c|}
    \toprule
    \textbf{Thresholds}
    &  30ms & 60ms & 90ms & 30ms & 60ms & 90ms \\
    \cmidrule(l){1-7}
    \textbf{Models} 
    & \multicolumn{3}{c|}{\textbf{Center Clipping}} 
    & \multicolumn{3}{c|}{\textbf{Peak Clipping}} \\
    \midrule
    Airport & 74.3 & 59.1 & 43 & 60.8 & 71.7 & 76.3 \\
    Babble & 73.2 & 59.3 & 42.2 & 57.2 & 69.7 & 73.7 \\
    Exhibition & 74.6 & 58.8 & 40.2 & 65.6 & 74.7 & 77.7\\
    Pink & 75.2 & 63.8 & 51 & 59.9 & 69.2 & 73.5 \\
    Restaurant & 71.3 & 52.3 & 34.6 & 60 & 70 & 75.2 \\
    Station & 74.8 & 60.5 & 42.7 & 61.1 & 71.7 & 74.7\\
    Street & 73.8 & 55.4 & 39.2 & 64 & 74.3 & 76.4 \\
    Train & 72.5 & 59.1 & 44.9 & 61.2 & 72.4 & 75.7 \\
    White Gaussian & 73.1 & 59.4 & 47.9 & 60.6 & 69.7 & 72.6 \\
    Clean & 68.2 & 47.7 & 32.1 & 59.6 & 70 & 75.1 \\
    \end{tabular}
    \label{tab:sre_dataset}
\end{table}

\begin{figure}[H]
    \centering
    \includegraphics[width=0.48\textwidth]{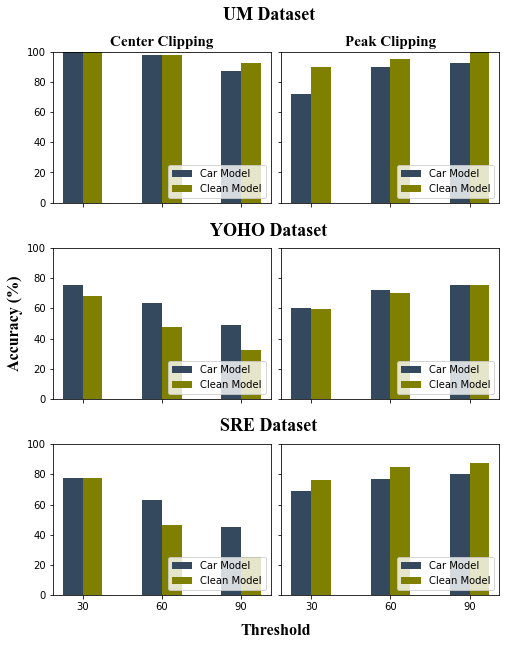}
    \caption{Performance of the car noise adapted model and the noise free model on distorted data}
    \label{fig:12}
\end{figure}
The result suggests that the proposed noise adapted speaker identification system is not only robust to noise and reverberation but also reluctant to distortion. The robustness becomes more obvious when noise is added along with other distortion factors as evident from the performance when both noise and reverberation were introduced.
\section{Discussion}
This study experimented the proposed noise adapted model based on cochleagram feature in the speaker identification task in clean and under noisy, reveberated and distorted conditions. After comparing the accuracy of clean and neurogram based models with the accuracy of the proposed model, it is obvious that the noise adapted model is more robust to noise when the signal-to-ratio is very low. Previously, the auditory nerve based neurogram feature was found to outform the baseline feature based (e.g., MFCC, FDLP and GFCC coefficients) SID systems that could provide consistent only for narrowband cases. The neurogram feature showed consistency for both narrowband and wideband cases. In this study, the noise adapted approach was found more consistent not only to wideband and narrowband noises but also for various types of noises from $-5$dB to clean conditions. 
\begin{figure}[H]
    \centering
    \includegraphics[width=0.48\textwidth]{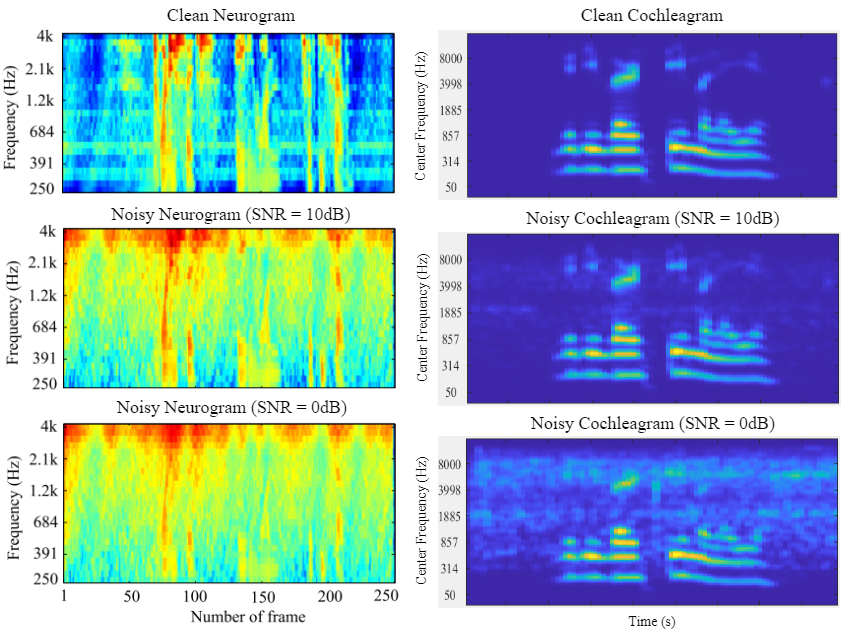}
    \caption{Illustration of the effect of noise on neurogram and cochleagram}
    \label{fig:13}
\end{figure}
The neurograms shown in fig. \ref{fig:13} were extracted from speech signal for $25$ characteristic frequencies logarithmically spaced between $250$ to $4000$Hz. For this the speech sample had to be resampled to $100$kHz to provide as input into the auditory nerve (AN) model for faithful replication of the frequency response properties of the AN model. However, increment of the sampling frequency also increases the computational cost. It can also be observed from the neurograms that the higher CFs are severely distorted with the increase in frequencies and the lower CFs are less distorted. On the contrary, the cochleagrams showed a finer resolution at low frequencies than neurograms. Besides, the cochleagrams were found to be comparatively less distorted with for addition of noises. The most important thing to mention is that the cochleagrams are retaining the finer frequency resolution after addition of noise. Another important feature of cochleagram which actually resembles how cochlea of inner ear in human analyzes sound is that it provides non-uniform time frequency resolution. The capacity to differentiate and separate distinct sounds perceptually is dependent on frequency selectivity in the inner ear, which is essential for hearing. This suggests that, the cochlear filter bandwidth may have a significant impact on speaker identification performance. However, delving into the details of each nonlinear occurrence identified at the peripheral level of auditory system's contribution to the SI task is beyond the scope of this research and might be undertaken as a future work. 

Evaluation of the noise adapted model's performance on various noises showed in fig.\ref{fig:6} revealed that the accuracy on text dependent UM dataset for all noises is $100\%$ from $10$dB to clean condition. In contrast, on text independent dataset although the accuracy of the noise adapted model from $-5$dB to $5$dB outperform the neurogram model, yet it did not improve with increase in SNR from $10$dB to clean condition. As for training clean and $-5$dB cochleagrams were used and the $-5$dB cochleagrams were rigorously distorted by noise, the model would require more data to learn more latent features from cochleagrams which could help the model to identify speaker more accurately and it became clear when the same model was trained using noisy data of two different SNR levels. In that case, the training accuracy was observed to increase from $85\%$ to $95\%$.

For clean reverberated data, the model trained using clean data outshone the noise adapted model. Training the models with more data might help to surpass the accuracy of the clean model. In addition, when reverberation as well as noise was introduced, the adapted models' accuracy increased $2-3$ times higher than the noise free model. Hence, it can inferred that the proposed model can achieve better accuracy under mismatched conditions where speech is distorted by several factors. As presented in fig. \ref{fig:12}, both the models' performance is approximately similar for center clipping and peak clipping. However, from the results presented earlier performance of the noise free model can easily be estimated if the accuracy is measured after addition of noise with distortion.

\section{Conclusion}
This study proposed a new technique to exploit the existing speech features to overcome the problem of speaker identification under noisy and mismatched environments and presented the way to make SID systems robust against various types of degradation of the input speech signals. To demonstrate the consistency under various conditions, the proposed model was treated with clean data, noisy data, reverberated data, noisy and reverberated and distorted data. The result showed no bias or dependency of the model on any noise rather a stable accuracy was observed irrespective of the narrowband as well as wideband cases across various noises. Besides, the approach showed that it is suitable for text dependent as well as text independent datasets. Although the model could be evaluated using speech corrupted with noise, reverberation and distortion all together and by addition of distortion and noise, it can be inferred from the previous results that the proposed model will certainly perform better. However, other existing features such as MFCC, CFCC and FDLP though they were found to be ineffective under noisy conditions yet the noise adapted approach can be adopted in future work as well as the neurogram feature can also be utilized.
% ACKNOWLEDGEMENT
\section*{Acknowledgment}
The authors would like to thank Bin Gao for helping with the matlab implementation of cochleagram model.

% BIBLIOGRAPHY
\bibliographystyle{IEEEtran}
\bibliography{References/manuscript.bib}

\end{document}